\documentclass[english,pra,final,twocolumn,superscriptaddress,showpacs]{revtex4-1}  
\pdfoutput=1
\usepackage[pdftex]{graphicx}    

\usepackage{babel}
\usepackage{textcomp}
\usepackage{amsmath}
\usepackage{amssymb}
\usepackage{amsfonts}
\usepackage{amscd}
\usepackage{amsbsy}         
\usepackage{bm}             
\usepackage{graphicx}       
\usepackage{dcolumn}        
\usepackage{latexsym}
\usepackage[latin1]{inputenc}   
\usepackage{float}          

\newcommand {\INLN} {Institut Non Lin\'{e}aire de Nice, CNRS and
Universit\'e Nice Sophia-Antipolis, 1361 route des Lucioles, 06560
Valbonne, France}

\newcommand {\at} {\mathrm{at}}
\newcommand {\scat} {\mathrm{sc}}
\newcommand {\kB} {k_\mathrm{B}}
\newcommand {\D} {\mathrm{D}}

\begin{document}
\title{Microscopic characterization of L\'{e}vy flights of light in atomic vapors}

\author{N. Mercadier}
\altaffiliation[Present address: ]{Saint-Gobain Recherche, 39 quai Lucien Lefranc, 93303 Aubervilliers, France}
\affiliation{\INLN}

\author{M. Chevrollier}
\affiliation{Departamento de F\'{i}sica, Universidade Federal da Para\'{i}ba, Jo\~{a}o Pessoa, PB, Brazil}

\author{W. Guerin}
\affiliation{\INLN}

\author{R. Kaiser}
\email[]{robin.kaiser@inln.cnrs.fr}
\affiliation{\INLN}

\date{\today}

\begin{abstract}
We investigate multiple scattering of near-resonant light in a Doppler-broadened atomic vapor. We experimentally characterize the length distribution of the steps between successive scattering events. The obtained power law is characteristic of a superdiffusive behavior, where rare but very long steps (L\'evy flights) dominate the transport properties.
\end{abstract}


\maketitle

\section{Introduction}

The diffusion model for the propagation of light has been used as early as 1922 by Compton to describe the transport of light in an atomic vapor \cite{Compton1922,Compton1923}. Soon afterwards, however, pioneering experiments by Zemansky on the decay of the fluorescence emitted by an initially excited mercury vapor have shown a deviation
from the prediction of such a diffusion model \cite{Zemansky1927}. Kenty realized that the frequency change during the scattering in a Doppler-broadened medium leads to lower excitation probabilities for photons with frequencies far from the center of the atomic resonance \cite{Kenty1932}. These photons can thus propagate over larger distances and escape a finite size system with increased probability. Soon afterwards, Holstein proposed an integro-differential equation allowing us to describe the transport of light taking into account the distribution of step lengths of the photons \cite{Holstein1947}. In the radiative transfer equation for the light propagation of photons at fixed frequency, this step-length distribution is an exponentially decreasing function, with well defined mean free path and higher moments. For the diffusion model to fail, a divergence of the second moment of the step-length distribution is required. Holstein showed that if the frequency of the photons inside the atomic vapor follows a Gaussian distribution (motivated by the Gaussian velocity distribution of the atoms) the step-length distribution of the photons has a divergent second moment, in line with the observations of Zemansky. A similar model was developed independently a few years later in the context of astrophysics to describe the radiative transfer in stellar atmospheres out of local thermal equilibrium \cite{Thomas1957,Jefferies1958,Thomas1960}.

If the step-length distribution of the photons $P(x)$ asymptotically follows a power law,
\begin{equation}
P(x)\longrightarrow \frac{1}{x^{\alpha}},\quad x\longrightarrow \infty  , \label{asympt}
\end{equation}
with $\alpha >3$, the variance of $P(x)$ is finite and the diffusion regime is normal. If $1< \alpha \leq 3$, the variance is not defined any more and the regime is said to be superdiffusive. It has been predicted \cite{Pereira2004} and demonstrated experimentally \cite{Mercadier2009} that the step-length distribution of light in resonant atomic vapors follows a power law with $\alpha <3$.

In this article, we report on the detailed experimental study of the length distribution of steps of quasi-resonant photons during their random walk in an atomic vapor. Particularly, we measure this distribution in three distinct regimes: (i) The single scattering regime, where photons with the same frequency originating from a laser are scattered once outside the laser beam volume, (ii) a double scattering regime, where photons scattered once at $90^{\circ}$ have a Doppler-broadened frequency spectrum before we measure their step-length distribution, and (iii) a multiple scattering limit, where photons have undergone many scattering events before their step-length distribution is measured. In the single scattering regime, we measure the step-length distribution and show that it still carries the memory of the frequency spectrum of the incident photons. We show that in the multiple scattering regime, on the contrary, the emission spectrum converges to a stable one, i.e., light ``thermalizes''. Therefore, the measurement of the step-length distribution after several scattering events allows us to characterize the transport properties.

The paper is organized as follows. In Sec. \ref{sec:Doppler}, we derive the expression of the step-length distribution for resonant photons scattered by Doppler-broadened atoms in the regime of complete frequency redistribution and show that its variance diverges, characterizing a superdiffusive transport of photons in the vapor. In Sec. \ref{sec:setup}, we detail the experimental protocol enabling us to measure the distance traveled by laser photons between their first and second scattering by atoms of a rubidium vapor and show that the obtained step-length distribution exhibits indeed an infinite variance. However, the deviation between the measurements and the predictions of our first model leads us to refine the latter, in Sec. \ref{sec:firststep}. We also illustrate the peculiar nature of the first scattering event, due to the memory that the photons keep of their initial frequency. We show in Sec. \ref{sec:multiple} that this correlation vanishes for a sufficiently large number of steps and that the corresponding step-length distribution is indeed characteristic of a superdiffusion regime. We conclude in Sec. \ref{sec:discussion}.

\section{Step-length distribution in a Doppler-broadened atomic vapor}\label{sec:Doppler}

\subsection{Frequency redistribution and broadening of the absorption profile by the Doppler effect}

A two-level atom with velocity \textbf{v}, illuminated by light of intensity well below its saturation intensity, elastically scatters the photons in its rest frame. Assuming that \textbf{v} is small compared to the speed of light, a photon of frequency $\omega$ in the laboratory frame and incident along the direction $\mathbf{e}_x$ has a frequency $\omega(1-\mathbf{v}\cdot\mathbf{e}_x/c)$ in the atomic rest frame. After scattering along the direction $\mathbf{e}'$, the frequency $\omega'$ of the photon as seen by an observer in the laboratory becomes
\begin{equation}
\omega' = \omega \left( 1-\frac{\mathbf{v}\cdot\mathbf{e}_x}{c} \right) \left( 1+\frac{\mathbf{v}\cdot\mathbf{e'}}{c} \right) \; .
\end{equation}
Thus, in a gas at non-zero temperature $T$, the Doppler effect contributes to the frequency redistribution of the incident light during a scattering process. Other effects, such as inelastic scattering by an isolated atom at rest or phenomena of energy exchange during inter-atomic collisions \cite{Takeo1957}, may also contribute to the frequency redistribution.

Since the atomic scattering cross section strongly varies with the frequency close to the atomic resonance, photons shifted from resonance by Doppler effect may travel in the medium a distance much longer than resonant photons. However, the extinction cross section for a moving atom is also shifted by the Doppler effect. The absorption profile $\Psi(\omega)$, inverse of the mean free path $\ell_\scat$ at the frequency $\omega$, is therefore determined by averaging over the atomic velocity distribution
\begin{equation}
\label{eq:mean_free_path}
\Psi(\omega) = \frac{1}{\ell_\scat(\omega)} =
n_{0}\int_{-\infty}^{+\infty} \sigma_\scat
\left[\omega\left(1-\frac{v_x}{c}\right)\right] \, P_\mathrm{M,1}(v_x) dv_x \; ,
\end{equation}
where $n_0$ is the atomic density, $P_\mathrm{M,1}$ is the Maxwell distribution of atomic velocities along a direction $x$, and $\sigma_\scat$ is the atomic scattering cross section at the frequency $\omega(1-v_x/c)$ of the light in the atomic rest frame.


From Beer-Lambert's law, the probability $P(x,\omega)$ that a photon of frequency $\omega$ travels a distance $x$ in the medium before it is scattered can be written
\begin{equation}\label{eq:Pxw}
P(x,\omega) = \frac{1}{\ell_\scat(\omega)} e^{- x/\ell_\scat(\omega)} \; .
\end{equation}

The distribution $P$ governing the length of a step is then obtained by averaging this result over the spectrum $\Theta(\omega)$ of the light emitted in the preceding scattering event,
\begin{equation}
\label{eq:P}
P(x)= \int_0^{+\infty} \Theta(\omega) P(x,\omega) d\omega \; .
\end{equation}
This distribution remains unchanged during the random walk of light, only if the emission spectrum $\Theta(\omega)$ is independent of the scattering event considered and, particularly, of the frequency of the photon during the previous step. This is the hypothesis of complete frequency redistribution (CFR), which is verified when some collisional broadening mechanisms dominate \cite{Molisch1998}. It is however incompatible with the description of Doppler broadening previously made.

\subsection{Simplification}

Even if the CFR hypothesis is \textit{a priori} not verified with Doppler broadening, it is still valuable to make this assumption in order to obtain analytical results. We suppose also that the emission profile is proportional to the absorption profile, which is the case in an infinite medium at thermodynamic equilibrium \cite{Holstein1947}. In the limit where the atomic scattering cross section is described by a Lorentzian of width $\Gamma$ much smaller than the Doppler width $\Delta\omega_\D$, the absorption profile tends to a Gaussian,
\begin{equation}\label{eq:Psi}
\Psi(\omega) =  \frac{1}{\ell_0} \frac{\pi \Gamma}{2} \frac{1}{\sqrt{2\pi} \, \Delta\omega_\mathrm{D}}
e^{-\frac{1}{2}\left(\frac{\omega-\omega_0}{\Delta\omega_\mathrm{D}}\right)^2}
\; ,
\end{equation}
where $\ell_0 = 1/[n_0 \sigma_\scat(\omega_0)]$ is the mean free path that a photon at resonance with the atomic transition of frequency $\omega_0$ would have if all the atoms were at rest (zero temperature). The width of the Doppler profile is
\begin{equation}\label{eq:dopplerwidth}
\Delta \omega_\D = \frac{\omega_0}{c} \sqrt{\frac{\kB T}{m}} \; ,
\end{equation}
where $\kB$ is the Boltzmann constant, $T$ is the temperature and $m$ is the atomic mass.
Moreover, by hypothesis,
\begin{equation}
\label{eq:spectre_gaussien}
\Theta(\omega) = \frac{1}{\sqrt{2\pi} \, \Delta\omega_\mathrm{D}}
e^{-\frac{1}{2}\left(\frac{\omega-\omega_0}{\Delta\omega_\mathrm{D}}\right)^2}
\; .
\end{equation}

\begin{figure}[t]
   \centering
   \includegraphics{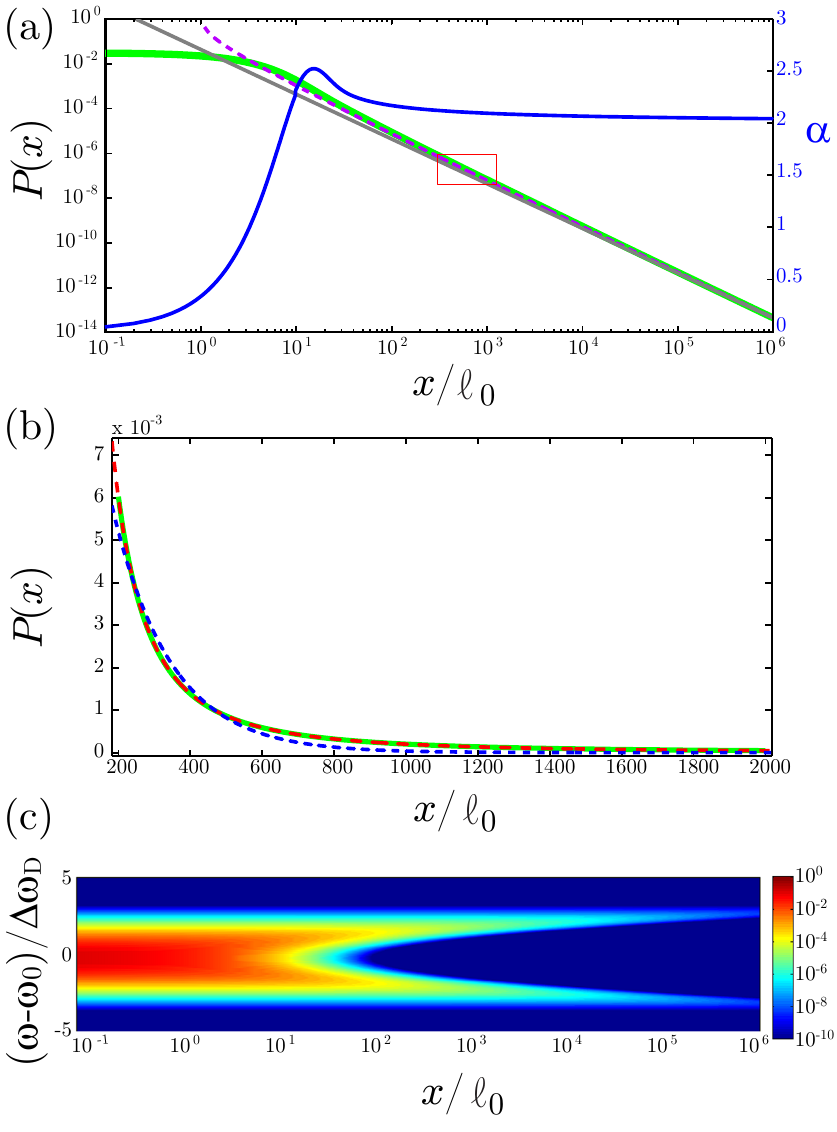}
  \caption{(Color online) (a) Step-length distribution (green, solid line), in logarithmic scale on both axes, estimated numerically from Eqs.~(\ref{eq:Pxw}-\ref{eq:spectre_gaussien}), for gaussian emission and absorption profiles. It is approximated by its asymptotic equivalent given by Eq. (\ref{eq:asymptote}) (purple, dashed line) and by a power law $1/x^2$ (gray, solid line). The blue, solid line shows the local slope of the step-length distribution, which gives the coefficient $\alpha$ (right axis) of the power law $x^{-\alpha}$ approaching the best the $P$ distribution. The red frame represents the typical window of experimentally reachable parameters.
  (b) The same step-length distribution (green, solid line) in this experimentally reachable window and in linear scale. It is compared to a fit by an exponential law (blue, dashed line) and by a power law (red, dashed line) given by Eq. (\ref{asympt}) with $\alpha=2.112\pm0.001$. We notice the good agreement between the power-law model and the numerically calculated distribution.
  (c) Evolution of the spectral profile of the light as a function of the distance $x$ in the atomic vapor. Photons in the wings of the atomic line propagate over a longer distance than photons at the line center.}
  \label{figureV1}
\end{figure}

By inserting the emission and absorption profiles [Eqs.~(\ref{eq:Psi}) and (\ref{eq:spectre_gaussien})] into Eqs.~(\ref{eq:Pxw}) and (\ref{eq:P}), we obtain the step-length distribution [Fig. \ref{figureV1}], whose  asymptotic behavior obeys \cite{Holstein1947,Pereira2004}
\begin{equation}
\label{eq:asymptote}
P(x) \sim \frac{1}{x^2 \sqrt{\ln(x/\ell_0)}} \; .
\end{equation}
It has a divergent second moment $\langle x^2 \rangle$, so that, in this framework, the transport of light in an atomic vapor falls under the scope of abnormal diffusion. One can show that the obtained distribution verifies a generalized central limit theorem \cite{Gnedenko1954}. Neglecting time-dependent aspects, we can therefore describe the resulting transport of light in terms of L\'{e}vy flights. Note that here, $P(x) = O(x^{-2})$, so that the mean free path is always finite, in contrast to the case of Lorentzian absorption and emission profile, where even the mean free path becomes infinite \cite{Pereira2004,footnote0}. Note also that the asymptotic behavior (\ref{eq:asymptote}) is close to a power law. In particular, in the range of $x/\ell_0$ accessible to our experiments, there is no noticeable difference between the numerically computed distribution and a power law $P(x)=x^{-\alpha}$, with $\alpha=2.112\pm0.001$ (confidence interval 95\%) [Fig.~\ref{figureV1}]. In the following, we will fit the computed and measured distributions by such a power law with an adjustable coefficient $\alpha$.

\section{Measurement of the step-length distribution}\label{sec:setup}

\subsection{Experimental setup}

A simple, yet precisely designed experimental set-up \cite{Mercadier_PhD} allows us to directly measure the distribution $P(x)$ of the length of the steps of photons between two scattering events in an atomic vapor. A beam with a power of $0.5$~mW, a 2~mm-waist and a spectral width smaller than 1~MHz is sent to a first cylindrical cell (18~mm diameter, 20~mm long) at room temperature $T_0=20^{\circ}$C containing a natural mixture of rubidium isotopes  ($^{85}$Rb : $72.17\%$, $^{87}$Rb : $27.83\%$) \cite{Stecka}. The beam is produced by an extended-cavity diode laser, stabilized to the transition $F=3 \rightarrow F'=4$ of the rubidium 85 D$_2$ line.

The atomic density in the cell is determined by the saturated vapor pressure. Fine adjustment of the cell temperature allows us to significantly change the atomic density and, consequently, the scattering mean free path for resonant light $\ell(\omega_0)$, while keeping the Doppler width $\Delta \omega_\D$ almost unchanged. At $T_0=20^{\circ}$C, the atomic density in the first cell is $\sim 9 \times 10^{15}$~m$^{-3}$, and the mean free path of the resonant light is $\sim 70$~mm. The optical depth in this cell is therefore at most $0.3$ in any direction, which ensures that photons undergo at most one scattering event, with a position uncertainty much smaller than $\ell(\omega_0)$.

\begin{figure}[t]
   \centering
   \includegraphics{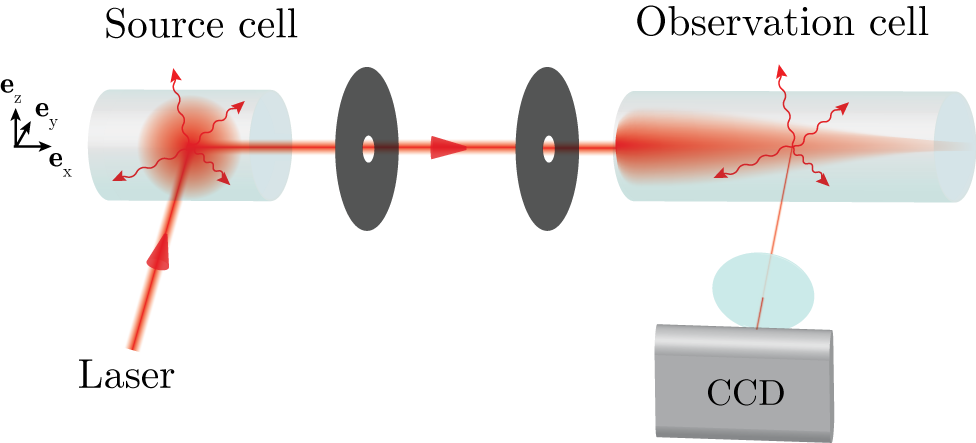}
  \caption{(Color online) Experimental set-up. A laser beam illuminates along $\mathbf{e}_y$ a first ``source cell" containing a rubidium vapor where incident photons are scattered at most once. A beam of scattered light propagating in the orthogonal direction $\mathbf{e}_x$ is selected by two diaphragms and sent to a second rubidium cell, called observation cell. The image of the fluorescence radiation from the second cell is projected onto a cooled CCD camera. The intensity detected along the axis of the incident beam gives the step-length distribution $P(x)$.}
  \label{figureV3}
\end{figure}

From the radiation scattered in this first cell, two iris diaphragms separated by 12~cm select a 2~mm-diameter beam propagating in a direction orthogonal to the initial laser [Fig.~\ref{figureV3}]. This beam crosses a second cylindrical rubidium cell (25~mm diameter, 75~mm long), with an angle of about $10^{\circ}$ with respect to the cell axis in order to prevent reflections at the cell sides from superimposing on the incident beam. This second cell is mounted on a heating plate that allows us to adjust its temperature. We can adjust the mean free path of the resonant light, from 70~mm at $20^{\circ}$C to 5~mm at $47^{\circ}$C.

We form the image of the fluorescence from the observation cell on a cooled CCD camera using a single lens (focal length 50~mm, diameter 25~mm), placed about 50~cm from the observation cell. While the small solid angle of detection limits the flux available for detection, it  allows us to suppress any vignetting effect and to obtain an excellent image flatness \cite{footnote1}. The CCD camera, an Apog\'{e}e AP2P, is equipped with a KAF-1602E-1 Kodak sensor with $1024 \times 1056$ pixels, a quantum efficiency of $60\%$ at $780$~mm  and a 14-bit dynamics. Its temperature is maintained at about $-6^{\circ}$C by a Peltier cooler. Each pixel on the recorded image corresponds to a size of 53.8 $\mu$m in the observation cell.

\begin{figure}[t]
   \centering
  \includegraphics{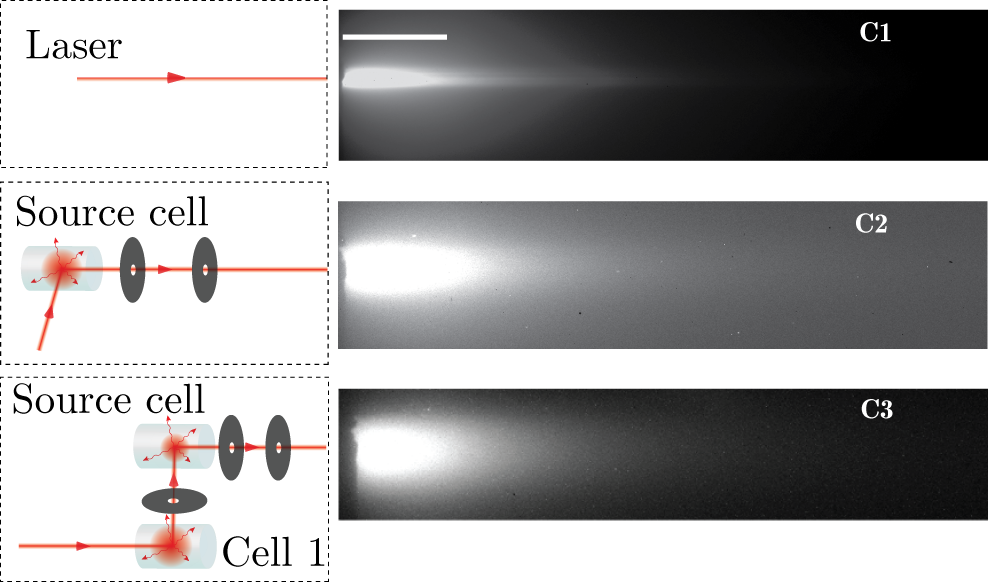}
  \caption{(Color online) Images obtained for the different experimental configurations. (C1) A resonant laser (locked to the $F=3 \rightarrow F'=4$ transition) directly illuminates the observation cell. (C2) The incident beam comes from the single scattering of a laser in a first source cell [Fig.~\ref{figureV3}]. (C3) Multiply-scattered photons in a first cell are used instead of the laser to illuminate the previous ``source cell''. For these three images, the temperature of the observation cell is $\sim 41^{\circ}$C, which corresponds to an atomic density in the cell of $5 \times 10^{16}$m~$^{-3}$ and a mean free path of $12$~mm for resonant light (white bar). Images C1 and C2 are obtained after a 30-minutes exposure and a dark image subtraction. Image C3 is obtained from six raw images with 5h exposure each and four corresponding dark images, according to the procedure detailed in section  \ref{sec:treatment}. The small dark region on the left of each image is due to a mask, which hides possible direct reflections of the incident beam off the glass of the observation cell.}
  \label{figureV5}
\end{figure}

In fact, three different experimental arrangements are used [Fig. \ref{figureV5}]. By sending directly a resonant laser in the observation cell, we can observe its exponential attenuation, characteristic of the Beer-Lambert law, and deduce the mean free path of the resonant light (C1 configuration). The atomic density in the cell is then obtained from Eq. (\ref{eq:mean_free_path}). The second configuration is the one detailed above [Fig.~\ref{figureV3}]: Photons incident in the observation cell have previously been scattered once in the source cell (C2 configuration). In the third experimental configuration (C3), the laser illuminates a first rubidium cell (18 mm~diameter, 30~mm long) heated to $36^{\circ}$C, whose optical depth along the long axis is $2.5$. In this cell, photons are scattered on average about 4 times before escaping. Photons propagating in the direction perpendicular to the exciting laser beam are selected by an iris diaphragm and sent to the source cell of configuration C2. In configuration C3, the available flux for detection by the CCD camera is extremely weak, on the order of 0.5 photon per pixel per hour in the image areas the most distant from the source [Fig.~\ref{figureV6}].

If the atomic density in the observation cell is weak enough for the photon to be in the single scattering regime, the detected intensity on the beam axis at the point of abscissa $x$ is proportional to the number of photons scattered after a step of length $x$ traveled in the observation cell. A section of the intensity detected along the axis of the incident beam [Fig. \ref{figureV8}] then gives the step-length distribution $P(x)$. In practice, if the atomic density in the cell is too weak, the attenuation of the ballistic beam is not sufficient to characterize its behavior (the width of the observation window represented in Fig. \ref{figureV1} decreases and its center is shifted to the low values of $x/\ell_0$). It is therefore necessary to increase the atomic density, with the drawback that the contribution of multiple scattering to the detected signal may not be negligible anymore. There is thus a tradeoff to find for the temperature of the second cell, and multiply-scattered light has to be taken into account in the image analysis.

\subsection{Image analysis}

\subsubsection{Noise reduction}\label{sec:treatment}

The image analysis uses a preprocessing stage aiming at removing biases and reproducible noise related to the imaging and the electronics of the camera, but also some non-reproducible artifacts. These procedures are particularly delicate for the images obtained in the last experimental configuration (with three cells) in conditions of extremely weak fluxes.


The offset due to charges accumulated in a pixel during the exposure time or added during the readout is efficiently corrected by subtracting from the raw signal a ``dark'' image, taken with the same exposure time and with the laser frequency shifted out of the Doppler absorption bands. Possible stray reflections are also eliminated with this procedure.

\begin{figure}[t]
   \centering
    \includegraphics{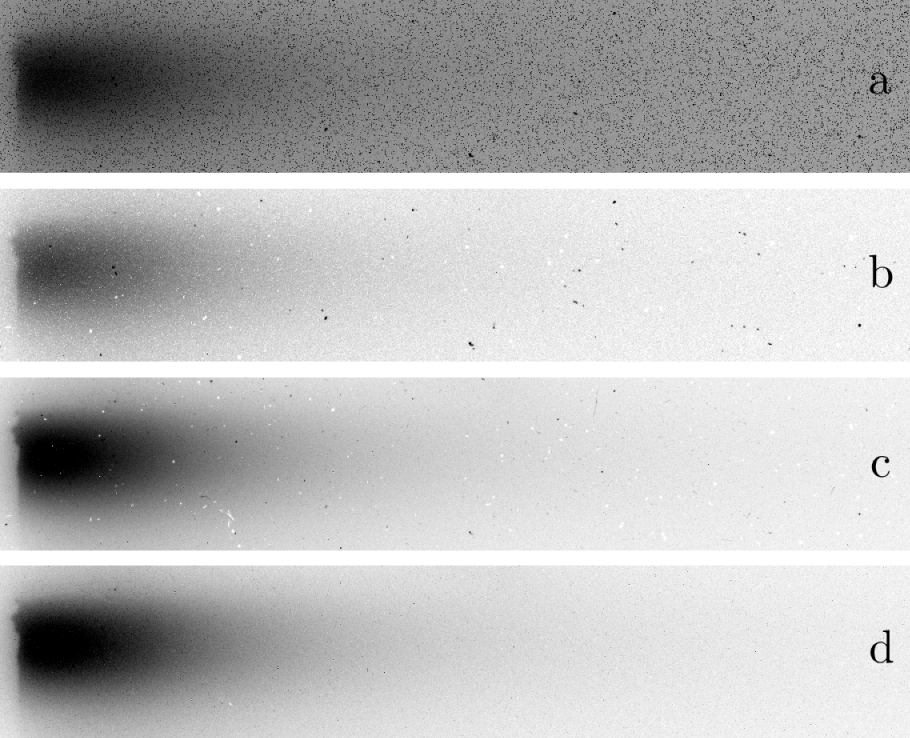}
  \caption{Images obtained for weak flux conditions (C3 configuration) via different processing. Gray levels have been inverted for better visibility. (a) Raw image, obtained with a 5h-exposure time. (b) After subtraction of a dark image. (c) Average of 6 raw images (30h total exposure), from which the average of four dark images has been subtracted. (d) Signal obtained by median compositing of 6 raw images, followed by subtraction of a median dark image.}
  \label{figureV6}
\end{figure}

Fluctuating noises cannot be filtered out via such image processing. At room temperature, the most significant of them is the thermal noise. Its impact is limited by a Peltier cooler, which keeps the sensor temperature at $-6^{\circ}$C. The noise related to sources present in the system is here essentially due to the photon shot-noise and to fluctuations of the number of detected electrons due to the direct impact of cosmic rays on the CCD sensor \cite{Groom2002}. Cosmic rays create locally important variations of the detected intensity, but the number of visible impacts on an image after a few hour exposure remains small. Thus, if several images are taken, it is unlikely that several of them will have impacts of cosmic rays on the same pixels.
We can therefore obtain an image clean from these impacts by taking for each pixel the median value of the intensity [Fig. \ref{figureV6}]. Although the suppression of thermal noise is less efficient than using all images for averaging, the obtained signal to noise ratio (SNR) is sufficient.

\subsubsection{Correction for multiple scattering}

If we neglect multiple scattering in the observation cell, the step-length distribution is simply given by the intensity along the incident-beam axis. In practice, we add the signals obtained in 30 (experiment with two cells) or 60 (three-cell configuration) lines of the CCD sensor (1.6 to 3.2 mm in the cell), and we average over the same number of pixels in the $x$ direction. This summation improves the signal-to-noise ratio (by a factor of 60 for a value of the intensity obtained by average over a $60\times 60$-pixel square). 

\begin{figure}[t]
   \centering
   \includegraphics{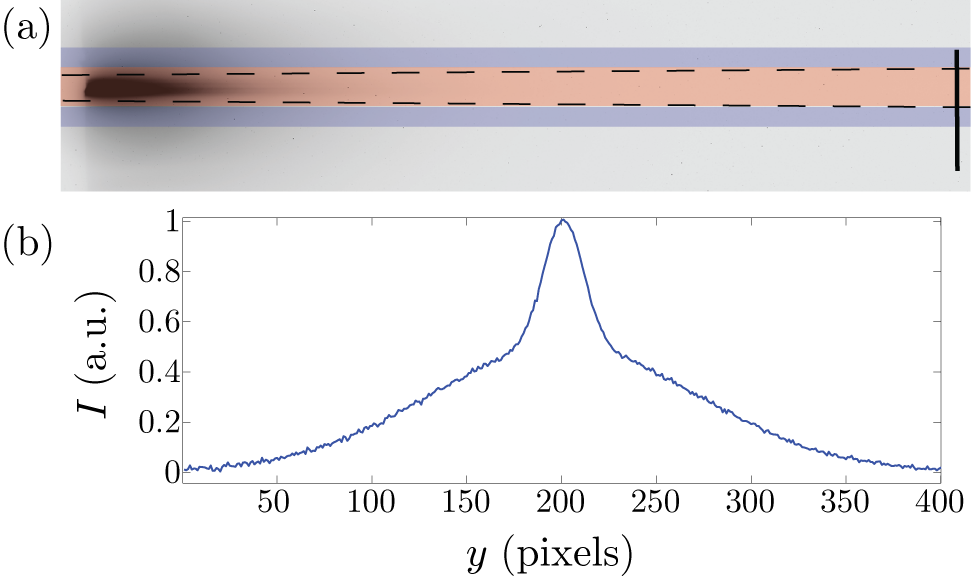}
  \caption{(Color online) (a) Fluorescence in the observation cell (inverted colors, C1 configuration). The intensity on and off axis is obtained through averaging over 60 lines (red and blue rectangles, respectively). The dashed lines represent the maximum width of the ballistic beam, as defined by the iris diaphragms. The vertical bar on the right represents the mean free path of resonant photons. (b) Vertical section of the image, where a central peak appears, related to the single scattering of the incident laser, and a larger structure (pedestal) due to multiple scattering.}
  \label{figureV8}
\end{figure}

\begin{figure}[b]
   \centering
   \includegraphics{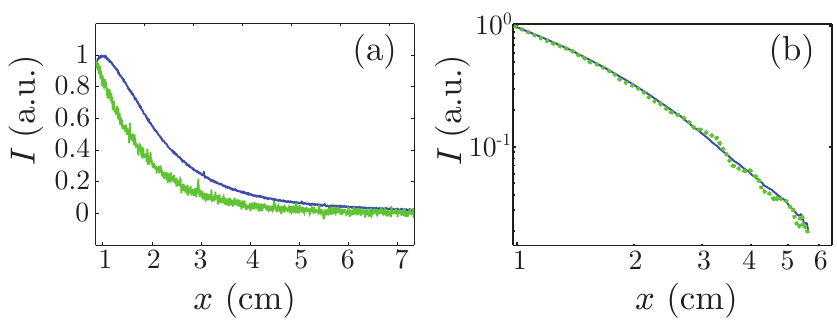}
  \caption{(Color online) (a) total intensity on the ballistic-beam axis (dark blue) in C2 configuration, and estimate of $P(x)$ (green) after correction for multiple scattering [Eq.~(\ref{eq:multscat})]. (b) Validation of the correction procedure by Monte-Carlo simulations: the on-axis intensity due to ballistic photons scattered only once (solid blue) is in very good agreement with the signal resulting from the subtraction of the off-axis intensity from the on-axis one (dotted green).}
  \label{figureV9}
\end{figure}

However, the intensity measured \emph{off-axis} of the ballistic beam is not zero, indicating that multiple scattering can not be neglected and thus affects the measurement of the intensity on the ballistic-beam axis. Indeed, at $41^{\circ}$C, the mean free path of the resonant photons is 12~mm and is comparable to the radius of the observation cell ($12.5$~mm). However, the effect of multiple scattering can be corrected. Indeed, the intensity measured along the ballistic-beam axis can be written in the form
\begin{equation}
I(x,0) = I_1(x,0) +I_{n \geq 2}(x,0) \; ,
\end{equation}
where $I_1$ is the intensity due to single scattering of ballistic photons and $I_{n \geq 2}$ is the intensity due to multiple scattering. Slightly off-axis, only this later contribution remains. If we assume a smooth variation of $I_{n \geq 2}(x,d)$ on a distance $d$ small compared to the mean free path of resonant photons, i.e., $I_{n \geq 2}(x,d) \simeq I_{n \geq 2}(x,0)$ for $d \ll \ell(\omega_0)$, we can use the off-axis measurement to subtract the multiple-scattering contribution from the intensity measured on axis.
Therefore,
\begin{equation}\label{eq:multscat}
P(x) \propto I(x,0)-I(x,d) \; .
\end{equation}
This correction process is illustrated in Fig. \ref{figureV8}. We have performed Monte-Carlo simulations to validate this correction procedure, see Fig. \ref{figureV9} and Ref. \cite{Chevrollier2010}. Also, the shape of the distribution $P(x)$ estimated through this method remains unchanged for measurements carried out with different atomic densities in the observation cell, and therefore for different amounts of multiple scattering (inset of Fig. \ref{figureV10}).


\section{Distribution of the first-step length}\label{sec:firststep}

\begin{figure}[t]
   \centering
   \includegraphics{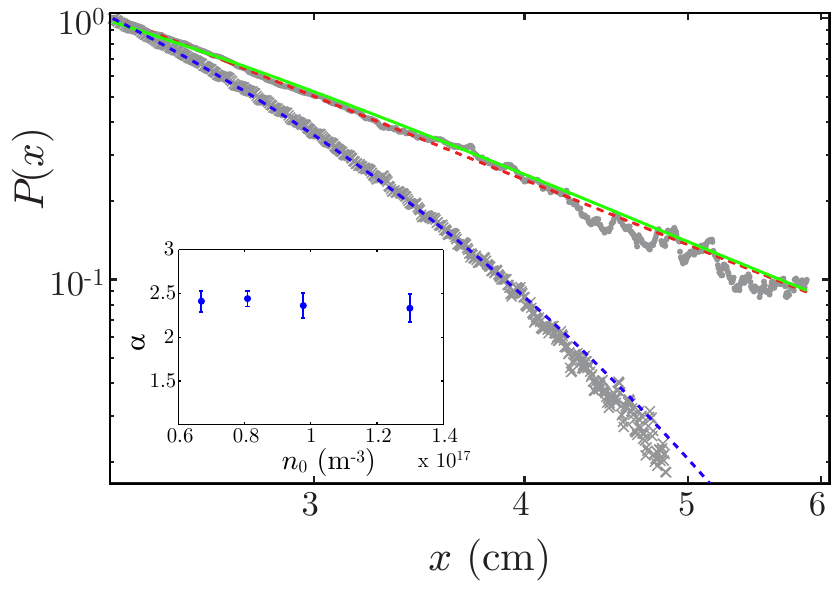}
  \caption{(Color online) Step-length distribution $P(x)$ in log-log scale. For a resonant and quasi-monochromatic incident laser (grey crosses), this distribution is well described by a decaying exponential (blue dashed line) corresponding to Beer's law. For an incident radiation coming from a first scattering in the source cell (C2 configuration), $P(x)$ exhibits a linear behavior in log-log scale, characteristic of a power law (grey points). A fit yields $P(x) \sim x^{-\alpha}$, with $\alpha = 2.41 \pm 0.12$ (red dashed line). This result is in very good agreement with the computed step-length distribution (green, solid line), see text. Inset: $\alpha$ coefficient measured for different atomic densities $n_0$ in the observation cell. The result is almost constant, demonstrating the effectiveness of the correction process for multiple scattering.}
  \label{figureV10}
\end{figure}

From images recorded in the two-cell configuration (C2), we obtain the step length distribution of the first step for photons scattered at $90^{\circ}$ in the source cell. The corresponding result is presented in Fig. \ref{figureV10}. A linear fit of the distribution $P(x)$ plotted in log-log scale shows that, in the experimentally reachable window and for sufficiently large $x$, $P(x)$ behaves as a power law,
\begin{equation}
P(x) \sim x^{-\alpha}, \quad \text{with } \alpha = 2.41 \pm 0.12 \; .
\end{equation}
Thus, with $\alpha <3$, the measured distribution has a diverging second moment. Consequently, standard diffusion can not properly describe the transport of light in an atomic vapor.

If the complete frequency redistribution hypothesis were verified, the measured law would govern the lengths of all steps and this result would allow the classification of the light transport as L\'{e}vy flights.
In the case of Doppler broadening, photon frequencies are redistributed around their initial frequency, leading to a memory effect incompatible \textit{a priori} with the hypothesis of complete frequency redistribution. Furthermore, the measured value of $P(x)$ is significantly different from the theoretical prediction $\alpha = 2.11$ expected in the experimentally reachable window (see Fig. \ref{figureV1}) made for Gaussian emission and absorption profiles. We thus have to take into account the precise emission and absorption spectra to correctly describe the experiment.

To calculate the length of the first step $P_1$ from Eq.~(\ref{eq:P}), we need to determine the absorption profile $\Psi$ of the rubidium vapor, as well as the spectrum $\Theta_{1,90^{\circ}}$ of the light scattered along $\mathbf{e}_x$, at $90^{\circ}$ of the direction $\mathbf{e}_y$ of the incident laser in the source cell,
\begin{equation}
\label{eq:P_2}
P_1(x)= \int_0^{+\infty} d\omega \;\Theta_{1,90^{\circ}} (\omega) \Psi(\omega) e^{- \Psi(\omega) x} \; .
\end{equation}
The absorption profile is given by Eq. (\ref{eq:mean_free_path}) and is independent of the number of previous scattering events.  The emission spectrum however evolves and, for a first scattering event, is given by
\begin{equation}\label{eq:spectra1}
\begin{split}
\Theta_{1,90^{\circ}}(\omega)  \propto
& \int_{0}^{+\infty} d\omega'
\int_{-\infty}^{+\infty}dv_x \int_{-\infty}^{+\infty}dv_y \,
\Theta_0 (\omega') \\
& \times \sigma_\scat \left[ \omega'\left(1-\frac{v_y}{c}\right) \right] \,
P_\mathrm{M,2}(v_x,v_y) \\
& \times \delta\left[\omega -
\omega'\left(1-\frac{v_y}{c}\right)\left(1+\frac{v_x}{c}\right)\right],
\end{split}
\end{equation}
where $\Theta_0(\omega')$ is the spectrum of the incident laser. The integral over $v_y$ yields the probability that an atom absorbs a photon of frequency $\omega_\at = \omega' (1 - v_y/c)$ in its rest frame. Knowing that a photon of frequency $\omega_\at$ has been absorbed, the integral over $v_x$ yields the probability that it is reemitted along $\mathbf{e}_x$ with frequency $\omega=\omega_\at (1+v_x/c)$. $P_\mathrm{M,2}$ is the Maxwell-Boltzmann distribution of velocities along two directions.
Finally, the Dirac distribution in the integral expresses energy conservation during the scattering process in the atomic rest frame.
\ \\

\noindent\textit{Case of an infinitely-narrow transition}\\
For a monochromatic incident laser, and in the limit where the width $\Gamma$ of the atomic transition is negligible compared to the Doppler width, the emission spectrum takes the gaussian shape given by Eq. (\ref{eq:spectre_gaussien}). The simple model assuming identical Gaussian emission and absorption profiles is thus relatively relevant. In this model, for the distances $x$ experimentally reachable, the distribution $P_1(x)$ is well described with a power law with parameter
\begin{equation}
\alpha = 2.112 \pm 0.001 \; .
\end{equation}
\ \\

\begin{figure}[t]
   \centering
   \includegraphics{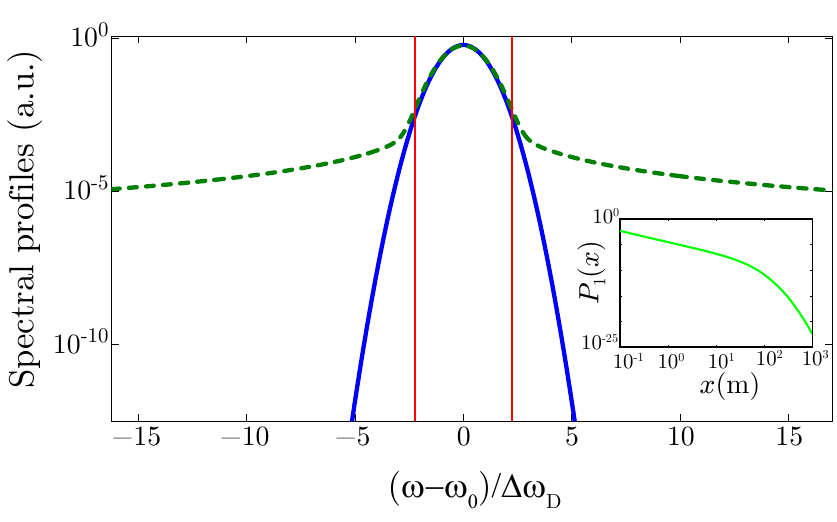}
  \caption{(Color online) Emission $\Theta_{1,90^{\circ}}$ (solid blue) and absorption (dashed green) spectra
for two-level atoms of natural width $\Gamma/2\pi = 6.066$~MHz, at temperature $T = 41^{\circ}\mathrm{C}$ (semi-logarithmic scale). The red lines define the detunings at which the mean free path is smaller than 1~m. In this frequency range, emission and absorption spectra are similar and close to a gaussian curve. Inset: Step-length distribution computed with these profiles, in log-log scale. A truncation at long range is clearly observable.}
  \label{figureV11}
\end{figure}

\noindent\textit{Impact of the natural width of the transition}\\
For two-level atoms with a transition of finite natural width $\Gamma$, the scattering cross section is a lorentzian function with a full width at half maximum $\Gamma$. For the D$_2$ line of rubidium, we have $\Gamma/2 \pi = 6.066$~MHz. In this case, the distribution $P_1(x)$ can be numerically estimated from Eqs. (\ref{eq:mean_free_path},\ref{eq:P_2},\ref{eq:spectra1}), taking also into account the Lorentzian shape of the spectrum of the initial laser (full width at half maximum $\sim 1$~MHz). A linear fit in log-log scale gives then
\begin{equation}
\alpha = 2.27 \pm 0.04 \; .
\end{equation}
The emission profile, visible in Fig. \ref{figureV11}, keeps a shape close to a Gaussian one. On the other hand, the absorption spectrum is a Voigt profile, convolution of the atomic velocity Gaussian distribution and the Lorentzian cross section \cite{F.2010}. Light strongly detuned from the atomic resonance is absorbed more efficiently than in the pure Doppler case. In the experimentally-accessible range, this results in an increase of the parameter $\alpha$ describing the decay of the step-length distribution $P_1$. For larger $x$, we observe a ``truncation'', i.e., a sharp collapse of this distribution (inset of Fig. \ref{figureV11}).
\ \\

\noindent\textit{Impact of the temperature difference between the cells}\\
In our experimental setup, the temperature of the observation cell ($T_2=314 \mathrm{K}$) is slightly higher than the temperature of the source cell ($T_1 = 293 \mathrm{K}$). The absorption profile is consequently broader than the emission one. This results in a slight decrease of the $P_1$ distribution experimentally accessible. Taking into account this temperature difference, we numerically obtain
\begin{equation}
\alpha = 2.32 \pm 0.04 \; .
\end{equation}
\ \\

\noindent\textit{Impact of the multi-level structure of rubidium}\\
The multi-level structure of rubidium modifies the expression of the scattering cross section. Assuming that the population is equally distributed between the Zeeman sub-levels of the hyperfine ground states of rubidium, which is a good approximation for a room-temperature vapor, we get
\begin{widetext}
\begin{equation}
\label{eq:sigma_gen}
\sigma(\omega) = \frac{\lambda_0^2}{\pi} \sum_{F=2}^{3}
\sum_{F'=1}^{4} \frac{2F+1}{\sum_{F_1=2}^{3} (2F_1 +1)}
\times  \frac{S_{FF'}}{1+ 4 (\omega - \omega_{FF'})^2/\Gamma^2 }
\end{equation}
The coefficients $S_{FF'}$ are transition factors calculated from the Clebsch-Gordan and the Wigner 3-$j$ coefficients \cite{Brink1994,Stecka} and $\omega_{FF'}$ is the frequency of the transition between the hyperfine ground state $F$ and the excited hyperfine level $F'$.

Due to Doppler broadening, the laser excites all the transitions from the fundamental level $F=3$ to the excited levels. Part of the light may undergo inelastic Raman scattering, with the atom going to state $F=2$. The emission spectrum can then be written
\begin{align}
\Theta_1(\omega) \propto & \int_{0}^{+\infty} d\omega'
\int_{-\infty}^{+\infty}dv_x \int_{-\infty}^{+\infty}dv_y \,
\Theta_0(\omega')
\times \frac{\lambda_0^2}{\pi} \sum_{F_1=2}^{3}
\sum_{F_2=2}^{3} \sum_{F'=1}^{4}
\frac{2F_1+1}{\sum_{F=2}^{3} (2F +1)} \\
&\times \frac{S_{F_1F'}}{1+ 4 \left[\omega'(1-v_y/c) -
\omega_{F_1F'}\right]^2/\Gamma^2 }
\times \frac{S_{F_2 F'}}{\sum_{F=2}^{3} S_{F F'}}
\times~P_\mathrm{M,2}(v_x,v_y) \, \delta\left[\omega -
\omega'\left(1-\frac{v_y}{c}\right)\left(1+\frac{v_x}{c}\right)-\omega_{F_1F_2}\right] \; , \nonumber
\end{align}
\end{widetext}
where $F_1$ and $F_2$ are the initial and final states of the scattering process, $F'$ the intermediate excited state and $\omega_{F_1 F_2}$ is the hyperfine splitting between the two ground states. The second term of the second line corresponds to the Raman scattering probability. Due to this process, the atomic emission spectrum is the superposition of two quasi-Gaussian peaks, with a frequency separation $\omega_{F_1 F_2} \sim 3$~GHz.
The distribution $P_1$ computed from this expression is plotted in Fig.~\ref{figureV10}. The atomic density used in the expression of the absorption profile is determined in an independent experiment (measurement of the attenuation of an incident laser beam in C1 configuration) so that the only adjustable parameter of the model is the incident intensity. We notice that the agreement between the model and the experiment is excellent. A fit of the computed distribution yields
\begin{equation}
\alpha = 2.45 \pm 0.04 \; .
\end{equation}
\ \\

\noindent\textit{Summary}\\
The natural width of the scattering transition, the complex structure of energy levels of the scattering atoms and finally the temperature difference between the source and the observation cells in the experiment, lead to corrections in the shape of the step-length distribution $P_1$ relatively to the predictions of the model at the origin of our study. All tend to accelerate the decay of this distribution (see Table \ref{tab:alpha}). By including these effects, the model is in excellent agreement with the experimental observations.

\begin{table}[t]
\caption{Impact of the different effects affecting the value of the coefficient $\alpha$ of the power law $x^{-\alpha}$ that better models the first-step distribution $P_1(x)$, in the distance range considered $[2-6~\mathrm{cm}]$ and  at $T=41^{\circ}$C.}\center
\begin{tabular}{|l|l|}
\hline
\textbf{Situation} & {$\mathbf{\alpha}$} \\
\hline
2-level atoms & \\
\& infinitely narrow transition & $2.112 \pm 0.001$ (num.) \\
\hline
+ Natural width & $2.27 \pm 0.04$ (num.) \\
\hline
+ Temperature difference & $2.32 \pm 0.04$ (num.) \\
\hline
+ Multi-level structure & $2.45 \pm 0.04$ (num.) \\
~~ & $2.41 \pm 0.12$ (expt.) \\
\hline
\end{tabular}

\label{tab:alpha}
\end{table}

\section{Multiple scattering regime}\label{sec:multiple}

\subsection{Light thermalization}

The spectrum of light scattered by the atoms is likely to evolve during the diffusion process. Particularly, the frequency redistribution induced by the Doppler effect tends to broaden the spectrum at each step so that the measurement performed on the first step does not actually allow us to draw a rigorous conclusion about the system behavior in the multiple scattering regime. Light ``thermalization'', i.e., the convergence of the spectrum towards a stable one, should however occur in the case where the natural width of the transition is much smaller than the Doppler one $\Delta \omega_\D$. Indeed, a photon with frequency detuning $\Delta \gg \Delta \omega_\D$ can only be scattered by a very fast atom, with its velocity $v_1$ along the incident-light axis much larger than the width $\Delta v$ of the Maxwellian velocity distribution. Its detuning after scattering is essentially imposed by the atom velocity along the reemission axis, statistically smaller than $v_1$. The frequency of the scattered photon is therefore brought back closer to the atomic resonance.

In order to compute the evolution of the atomic emission spectrum with the number of scattering events, we consider a photon with initial frequency $\omega_0$, submitted to a random walk in an infinite gas, and we compute the probability $\Theta_n(\omega)$ that the photon frequency is $\omega$ after $n$ scattering events. $\Theta_n(\omega)$ is obtained by averaging the emission spectrum over all possible scattering angles \cite{Molisch1998,Alves-Pereira2007}. As we assume the medium to be infinite, the photon can not escape and is therefore always scattered, regardless of its frequency $\omega$ at a given instant. Using the joint redistribution function $R$, already averaged over the angles, and giving the probability that a photon of frequency $\omega'$ is scattered and reemitted at frequency $\omega$, we get
\begin{equation}
\label{eq:spectreEvolution}
\Theta_{n+1}(\omega) = \int_{-\infty}^{+\infty} \Theta_n(\omega')
\frac{R(\omega,\omega')}{\tilde{\Psi}(\omega')} \, d\omega' \; ,
\end{equation}
where $\tilde{\Psi}$ is the normalized absorption profile \cite{footnote2}.
In the simple case of two-level atoms with an infinitely narrow transition, the joint redistribution function is given by
\begin{equation}
\label{eq:redistribution}
R_{I}(\omega,\omega') = \frac{1}{2} \mathrm{erfc}(\overline{X}) \; ,
\end{equation}
where $\overline{X}=\max \left( \left|
\frac{\omega-\omega_0}{\sqrt{2}\Delta\omega_\mathrm{D}} \right|, \left|
\frac{\omega'-\omega_0}{\sqrt{2} \Delta\omega_\mathrm{D}} \right|
\right)$ \cite{Molisch1998}. The normalized, Gaussian absorption profile is given by Eq. (\ref{eq:spectre_gaussien}). As expected, an emission spectrum with the same shape remains unchanged through Eq. (\ref{eq:spectreEvolution}), which suggests that the measurement of the first-step length distribution previously performed gives a good estimate of the step-length distribution in the multiple scattering regime, due to the very specific selection of the first-scattering angle.

\begin{figure}[t]
   \centering
   \includegraphics{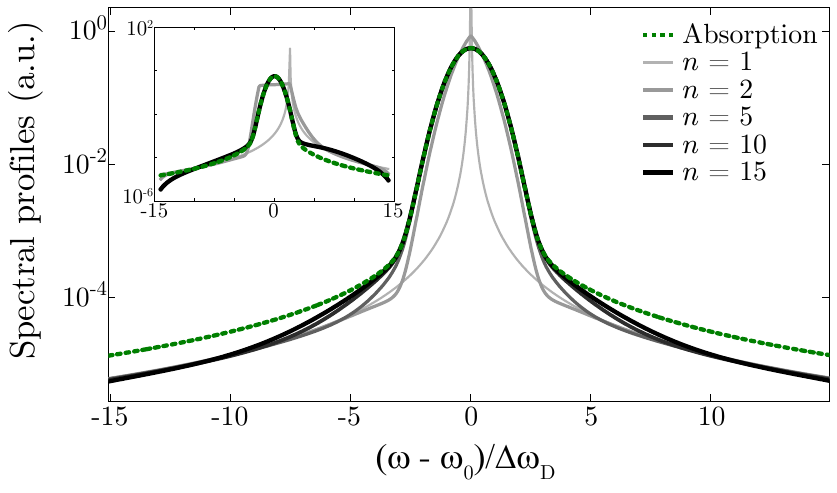}
  \caption{(Color online) Probability $\Theta_n(\omega)$ that the frequency of a photon is $\omega$, as a function of the number of steps $n$. The initial spectrum (light grey) is the Lorentzian spectrum of the laser used in the experiment, at resonance. We notice that the emission spectrum converges to the Voigt absorption profile (dashed green). Inset: Same emission spectra, for an initial laser detuned two Doppler widths from resonance (440~MHz).}
  \label{figureV12}
\end{figure}

If we take into account the finite width $\Gamma$ of the atomic transition, the redistribution function is
\begin{widetext}
\begin{equation}\label{eq:RII}
R_{II}(\omega,\omega') = \pi^{-\frac{3}{2}} \int_{\frac{1}{2}\mid
\frac{\omega-\omega'}{\sqrt{2}\Delta\omega_\mathrm{D}}
\mid}^{+\infty} e^{-u^2} \left[ \arctan
\left(\frac{\underline{X}+u}{a}\right) - \arctan
\left(\frac{\overline{X}-u}{a}\right) \right] du \; ,
\end{equation}
where $\underline{X} =\min\left(\left|
\frac{\omega-\omega_0}{\sqrt{2}\Delta\omega_\mathrm{D}} \right|, \left|
\frac{\omega'-\omega_0}{\sqrt{2}\Delta\omega_\mathrm{D}} \right|
\right)$ and $a=\Gamma/(\sqrt{8} \Delta \omega_\D)$ is the Voigt parameter \cite{footnote3}.
Figure \ref{figureV12} shows the evolution of the emission spectrum $\Theta_n$, numerically computed from Eqs. (\ref{eq:spectreEvolution}) and (\ref{eq:RII}) as a function of the number of steps. The figure highlights the convergence of the spectrum to the normalized Voigt absorption profile, faster for frequencies close to the line center than for frequencies in the wings of the distribution. Thus, after a few steps, the photon losses the memory of its initial frequency.

Finally, we can take into account the multi-level structure of rubidium by averaging the contributions of all hyperfine transitions, weighted by their transition factor. Including the inelastic Raman scattering, we obtain
\begin{equation}
R(\omega,\omega')=\sum_{F_1=2}^{3} \sum_{F_2=2}^{3}
\sum_{F'=1}^{4} \frac{2F_1+1}{\sum_{F=2}^{3} (2F +1) } \times
\frac{S_{F_1F'} S_{F_2 F'}}{\sum_{F=2}^{3} S_{FF'}} \times
R_{II}\left( \omega-\omega_{F_2F'},\omega'-\omega_{F_1F'}\right)
\; .
\end{equation}
\end{widetext}

We thus deduce the evolution of the emission spectrum via Eq. (\ref{eq:spectreEvolution}), and then the distribution $P_n$ governing the length of the step $n$ from Eqs. (\ref{eq:mean_free_path}), (\ref{eq:P}), and (\ref{eq:sigma_gen}). Following the same fitting procedure as before, we obtain the coefficient $\alpha$ of the power law that better models the distribution $P_n$ in the range of distance experimentally accessible. The evolution of this coefficient with the number of steps is reported in the inset of Fig. \ref{figureV13}. As the number of steps increases, the profile wings get broader and this coefficient decreases until it reaches, in the region experimentally accessible, a limit value close to the value obtained for two purely Doppler emission and absorption spectra.

\subsection{Measurement}

\begin{figure}[t]
   \centering
   \includegraphics{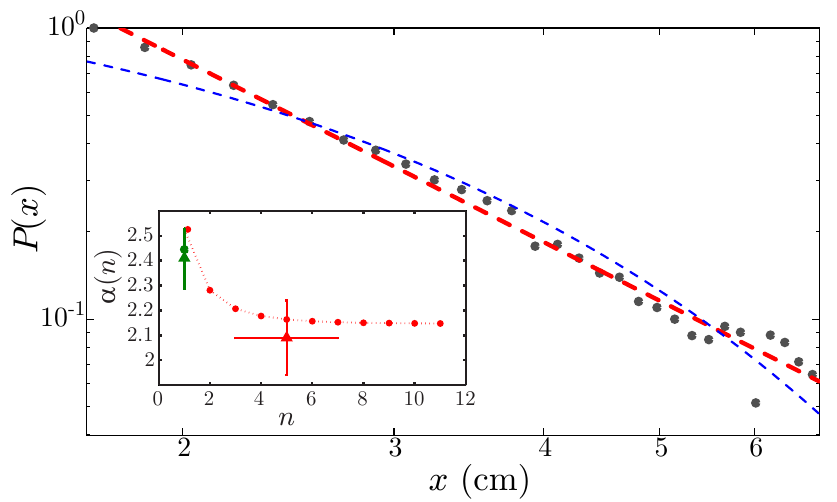}
  \caption{(Color online) Step-length distribution in multiple scattering regime (log-log scale), measured in the C3 configuration (black dots). The measurement is compared to a fit by an exponential law (dashed blue) and by a power law (dashed red) $P(x) \sim x^{-\alpha}$ with $\alpha = 2.08 \pm 0.13$. Inset: Evolution of the coefficient $\alpha$ as a function of the number of steps. The numerical result (red dots) is compared to the measurement in configurations C3 (red triangle) and C2 (green triangle). In the latter case, a model without angular averaging is more appropriate (green dot).}
  \label{figureV13}
\end{figure}

Due to the thermalization process, a single distribution $P(x)$ governs the step length during multiple scattering in an infinite medium. The experimental configuration C3 allows us to prepare photons having undergone several (typically 2 to 6) scattering events before we measure the step-length distribution, and thus to approach a measurement of the limit distribution $P(x)$ [Fig. \ref{figureV13}].
We measure
\begin{equation}
P(x) \sim x^{-\alpha} , \quad \textrm{with } \alpha = 2.08 \pm 0.13 \; .
\end{equation}
This result agrees very well with our numerical estimate of the distribution $P(x)$ of the step length in the multiple scattering regime in the distance range experimentally accessible.

\section{Conclusion}\label{sec:discussion}

The shape of the distribution governing the step length of the random walk of quasi-resonant light in a dilute alkali-metal atomic vapor at room temperature is entirely determined by the atomic emission and absorption spectra and evolves, with the emitted light spectrum, during the diffusion process. We have computed this evolution and we have shown that the step-length distribution converges to a limit law when the number of steps grows. We have also presented direct measurements of the distance traveled by light between two scattering events, first for the first step directly following the scattering of photons generated by a laser source, then for photons having experienced several scattering events and having thus lost memory of their initial frequency. The results are in very good agreement with our models taking into account, beyond the Doppler effect, the natural width of the transitions and the multilevel structure of rubidium 85. Above all, they show that, in the distance range experimentally accessible, the step-length distribution in the multiple scattering regime is described by a power law $P(x) \sim x^{-\alpha}$ with $\alpha < 3$. Thus, the second moment of this distribution diverges. Consequently transport of light in the stationary regime can be described by a L\'{e}vy-flight model.

A few important questions remain open, notably related to the temporal dynamics of abnormal diffusion of light in an atomic vapor or to the impact of correlations between steps. Most of known systems giving rise to a superdiffusive regime are spatially inhomogeneous, and the existence of these correlations is therefore inevitable. Their impact on the transport properties may however be very limited \cite{Barthelemy2010}. In atomic vapors, correlations can be suppressed, provided the regime of complete frequency redistribution is achieved. Addition of a buffer gas in rubidium cells may enable one to get close to this condition \cite{Ottinger1975}. It would affect the shape of the emission and absorption spectra, which could become Lorentzian, leading to an asymptotic decay of the step-length distribution too slow to ensure that the scattering mean free path remains defined.

Finally, atomic vapors constitute a model system, of simple experimental implementation, that may enable the characterization of different regimes of abnormal transport. This work shows that the microscopic element constituted by the step-length distribution is as experimentally accessible as macroscopic quantities characteristic of the transport, such as the diffuse transmission through a sample \cite{Barthelemy2008,Baudouin2013}.

\begin{acknowledgments}
We acknowledge fruitful discussions with R. Carminati, R. Pierrat and E. Pereira. The PhD grant of N.M. has been funded by DGA.
\end{acknowledgments}


\begin{thebibliography}{0}%
\makeatletter
\providecommand \@ifxundefined [1]{%
 \@ifx{#1\undefined}
}%
\providecommand \@ifnum [1]{%
 \ifnum #1\expandafter \@firstoftwo
 \else \expandafter \@secondoftwo
 \fi
}%
\providecommand \@ifx [1]{%
 \ifx #1\expandafter \@firstoftwo
 \else \expandafter \@secondoftwo
 \fi
}%
\providecommand \natexlab [1]{#1}%
\providecommand \enquote  [1]{``#1''}%
\providecommand \bibnamefont  [1]{#1}%
\providecommand \bibfnamefont [1]{#1}%
\providecommand \citenamefont [1]{#1}%
\providecommand \href@noop [0]{\@secondoftwo}%
\providecommand \href [0]{\begingroup \@sanitize@url \@href}%
\providecommand \@href[1]{\@@startlink{#1}\@@href}%
\providecommand \@@href[1]{\endgroup#1\@@endlink}%
\providecommand \@sanitize@url [0]{\catcode `\\12\catcode `\$12\catcode
  `\&12\catcode `\#12\catcode `\^12\catcode `\_12\catcode `\%12\relax}%
\providecommand \@@startlink[1]{}%
\providecommand \@@endlink[0]{}%
\providecommand \url  [0]{\begingroup\@sanitize@url \@url }%
\providecommand \@url [1]{\endgroup\@href {#1}{\urlprefix }}%
\providecommand \urlprefix  [0]{URL }%
\providecommand \Eprint [0]{\href }%
\providecommand \doibase [0]{http://dx.doi.org/}%
\providecommand \selectlanguage [0]{\@gobble}%
\providecommand \bibinfo  [0]{\@secondoftwo}%
\providecommand \bibfield  [0]{\@secondoftwo}%
\providecommand \translation [1]{[#1]}%
\providecommand \BibitemOpen [0]{}%
\providecommand \bibitemStop [0]{}%
\providecommand \bibitemNoStop [0]{.\EOS\space}%
\providecommand \EOS [0]{\spacefactor3000\relax}%
\providecommand \BibitemShut  [1]{\csname bibitem#1\endcsname}%
\let\auto@bib@innerbib\@empty
\end{thebibliography}%


\begin{thebibliography}{10}

\bibitem{Compton1922}
K. Compton, Phys. Rev. {\bf 20},  283  (1922).

\bibitem{Compton1923}
K. Compton, Philos. Mag. {\bf 45},  752  (1923).

\bibitem{Zemansky1927}
M. Zemansky, Phys. Rev. {\bf 29},  513  (1927).

\bibitem{Kenty1932}
C. Kenty, Phys. Rev. {\bf 42},  823  (1932).

\bibitem{Holstein1947}
T. Holstein, Phys. Rev. {\bf 72},  1212  (1947).

\bibitem{Thomas1957}
R.~N. Thomas, ApJ. {\bf 125},  260  (1957).

\bibitem{Jefferies1958}
J.~T. Jefferies and R.~N. Thomas, ApJ. {\bf 127},  667
  (1958).

\bibitem{Thomas1960}
R.~N. Thomas, ApJ {\bf 131},  429  (1960).

\bibitem{Pereira2004}
E. Pereira, J. Martinho, and M. Berberan-Santos, Phys. Rev. Lett. {\bf
  93},  120201  (2004).

\bibitem{Mercadier2009}
N. Mercadier, W. Guerin, M. Chevrollier, and R. Kaiser, Nat. Phys. {\bf 5},
   602  (2009).

\bibitem{Takeo1957}
S. Chen and M. Takeo, Rev. Mod. Phys. {\bf 29},  20  (1957).

\bibitem{Molisch1998}
A.~F. Molisch and B.~P. Oehry, {\em {Radiation Trapping in Atomic Vapours}}
  (Oxford University Press, New York, 1998), p.\ 510.

\bibitem{Gnedenko1954}
B.~V. Gnedenko, {\em {Limit Distributions for Sums of Independent Random
  Variables}} (Addison-Wesley, Cambridge, MA, 1954), p.\ 264.

\bibitem{footnote0}
The mean free path diverges also for Voigt emission and absorption profiles \cite{Pereira2004}.

\bibitem{Mercadier_PhD}
N. Mercadier, Ph.D. thesis, Universit\'e Nice Sophia Antipolis, 2011.

\bibitem{Stecka}
D.~A. Steck, {Rubidium 85 D Line Data, http://steck.us/alkalidata}, 2008.

\bibitem{footnote1}
In this configuration the solid angle-induced relative difference between the intensity
  at the center and at the edge of the image, for a perfectly homogeneous
  source, does not exceed $2 \times 10^{-4}$.

\bibitem{Groom2002}
D. Groom, Exp. Astron. {\bf 14},  45  (2002).

\bibitem{Chevrollier2010}
M. Chevrollier, N. Mercadier, W. Guerin, and R. Kaiser, Eur. Phys.
  J. D {\bf 58},  161  (2010).

\bibitem{F.2010}
F.~W.~J. Olver, D.~W. Lozier, R.~F. Boisvert, and C.~W. Clark, {\em {NIST
  Handbook of Mathematical Functions}} (Cambridge University Press, New York,
  2010), p.\ 968.

\bibitem{Brink1994}
D.~M. Brink and G.~R. Satchler, {\em {Angular Momentum}} (Oxford University
  Press, New York, 1994), p.\ 192.

\bibitem{Alves-Pereira2007}
R. Alves-Pereira, E.~J. Nunes-Pereira, J.~M.~G. Martinho, and M.~N.
  Berberan-Santos, J. Chem. Phys. {\bf 126},  154505  (2007).

\bibitem{footnote2}
$p(\omega, \omega') = \frac{R(\omega,\omega')}{\tilde{\Psi}(\omega')}$ gives
  the probability that a photon of frequency $\omega'$, having been absorbed,
  is reemitted with frequency $\omega$. We use $R$ as an intermediate quantity
  because this function is available in the literature for various broadening
  mechanisms.

\bibitem{footnote3}
This expression of the redistribution function $R_{II}$ is given in
  \cite{Molisch1998, Alves-Pereira2007}. Note that it is valid only in
  situations where the Doppler width is large compared to the natural width, so
  that it is not applicable to cold atoms.

\bibitem{Barthelemy2010}
P. Barthelemy, J. Bertolotti, K. Vynck, S. Lepri, and D. Wiersma, Phys. Rev. E
  {\bf 82},  011101  (2010).

\bibitem{Ottinger1975}
C. Ottinger, R. Scheps, G. York, and A. Gallagher, Phys. Rev. A {\bf 11},
  1815  (1975).

\bibitem{Barthelemy2008}
P. Barthelemy, J. Bertolotti, and D.~S. Wiersma, Nature (London) {\bf 453},  495
  (2008).

\bibitem{Baudouin2013}
Q. Baudouin, R. Pierrat, A. Eloy, E. Pereira, N. Mercadier, and R. Kaiser, in preparation.

\end{thebibliography}

\end{document}